# Fairness Testing:
# Testing Software for Discrimination

Sainyam Galhotra    Yuriy Brun    Alexandra Meliou
University of Massachusetts, Amherst
Amherst, Massachusetts 01003-9264, USA
{sainyam, brun, ameli}@cs.umass.edu

## ABSTRACT

This paper defines software fairness and discrimination and develops a testing-based method for measuring if and how much software discriminates, focusing on causality in discriminatory behavior. Evidence of software discrimination has been found in modern software systems that recommend criminal sentences, grant access to financial products, and determine who is allowed to participate in promotions. Our approach, Themis, generates efficient test suites to measure discrimination. Given a schema describing valid system inputs, Themis generates discrimination tests automatically and does not require an oracle. We evaluate Themis on 20 software systems, 12 of which come from prior work with explicit focus on avoiding discrimination. We find that (1) Themis is effective at discovering software discrimination, (2) state-of-the-art techniques for removing discrimination from algorithms fail in many situations, at times discriminating against as much as 98% of an input subdomain, (3) Themis optimizations are effective at producing efficient test suites for measuring discrimination, and (4) Themis is more efficient on systems that exhibit more discrimination. We thus demonstrate that fairness testing is a critical aspect of the software development cycle in domains with possible discrimination and provide initial tools for measuring software discrimination.

## 1 INTRODUCTION

Software has become ubiquitous in our society and the importance of its quality has increased. Today, automation, advances in machine learning, and the availability of vast amounts of data are leading to a shift in how software is used, enabling the software to make more autonomous decisions. Already, software makes decisions in what products we are led to buy [53], who gets a loan [62], self-driving car actions that may lead to property damage or human injury [32], medical diagnosis and treatment [74], and every stage of the criminal justice system including arraignment and sentencing that determine who goes to jail and who is set free [5, 28]. The importance of these decisions makes fairness and nondiscrimination in software as important as software quality.

Unfortunately, software fairness is undervalued and little attention is paid to it during the development lifecycle. Countless examples of unfair software have emerged. In 2016, Amazon.com, Inc. used software to determine the parts of the United States to which it would offer free same-day delivery. The software made decisions that prevented minority neighborhoods from participating in the program, often when every surrounding neighborhood was allowed to participate [36, 52]. Similarly, software is being used to compute risk-assessment scores for suspected criminals. These scores — an estimated probability that the person arrested for a crime is likely to commit another crime — are used to inform decisions about who can be set free at every stage of the criminal justice system process, from assigning bond amounts, to deciding guilt, to sentencing. Today, the U.S. Justice Department's National Institute of Corrections encourages the use of such assessment scores. In Arizona, Colorado, Delaware, Kentucky, Louisiana, Oklahoma, Virginia, Washington, and Wisconsin, these scores are given to judges during criminal sentencing. The Wisconsin Supreme Court recently ruled unanimously that the COMPAS computer program, which uses attributes including gender, can assist in sentencing defendants [28]. Despite the importance of these scores, the software is known to make mistakes. In forecasting who would reoffend, the software is "particularly likely to falsely flag black defendants as future criminals, wrongly labeling them this way at almost twice the rate as white defendants; white defendants were mislabeled as low risk more often than black defendants" [5]. Prior criminal history does not explain this difference: Controlling for criminal history, recidivism, age, and gender shows that the software predicts black defendants to be 77% more likely to be pegged as at higher risk of committing a future violent crime than white defendants [5]. Going forward, the importance of ensuring fairness in software will only increase. For example, "it's likely, and some say inevitable, that future AI-powered weapons will eventually be able to operate with complete autonomy, leading to a watershed moment in the history of warfare: For the first time, a collection of microchips and software will decide whether a human being lives or dies" [34]. In fact, in 2016, the U.S. Executive Office of the President identified bias in software as a major concern for civil rights [27]. And one of the ten principles and goals Satya Nadella, the CEO of Microsoft Co., has laid out for artificial intelligence is "AI must guard against bias, ensuring proper, and representative research so that the wrong heuristics cannot be used to discriminate" [61].

This paper defines causal software discrimination and proposes Themis, a software testing method for evaluating the fairness of software. Our definition captures causal relationships between inputs and outputs, and can, for example, detect when sentencing software behaves such that "changing only the applicant's race affects the software's sentence recommendations for 13% of possible applicants." Prior work on detecting discrimination has focused on measuring correlation or mutual information between inputs and outputs [79], discrepancies in the fractions of inputs that produce a given output [18, 19, 39–42, 87–89, 91], or discrepancies in output probability distributions [51]. These approaches do not capture causality and can miss discrimination that our causal approach detects, e.g., when the software discriminates negatively with respect to a group in one settings, but positively in another. Restricting the input space to real-world inputs [1] may similarly hide software discrimination that causal testing can reveal. Unlike prior work that requires manually written tests [79], Themis automatically generates test suites that measure discrimination. To the best of our knowledge, this work is the first to automatically generate test suites to measure causal discrimination in software.

Themis would be useful for companies and government agencies relying on software decisions. For example, Amazon.com, Inc. received strong negative publicity after its same-day delivery algorithm made racially biased decisions. Politicians and citizens in Massachusetts, New York, and Illinois demanded that the company offer same-day delivery service to minority neighborhoods in Boston, New York City, and Chicago, and the company was forced to reverse course within mere days [72, 73]. Surely, the company would have preferred to test its software for racial bias and to develop a strategy (e.g., fixing the software, manually reviewing and modifying racist decisions, or not using the software) prior to deploying it. Themis could have analyzed the software and detected the discrimination prior to deployment. Similarly, a government may need to set nondiscrimination requirements on software, and be able to evaluate if software satisfies those requirements before mandating it to be used in the justice system. In 2014, the U.S. Attorney General Eric Holder warned that steps need to be taken to prevent the risk-assessment scores injecting bias into the courts: "Although these measures were crafted with the best of intentions, I am concerned that they inadvertently undermine our efforts to ensure individualized and equal justice [and] they may exacerbate unwarranted and unjust disparities that are already far too common in our criminal justice system and in our society." [5]. As with software quality, testing is likely to be the best way to evaluate software fairness properties.

Unlike prior work, this paper defines discrimination as a causal relationship between an input and an output. As defined here, discrimination is not necessarily bad. For example, a software system designed to identify if a picture is of a cat *should* discriminate between cats and dogs. It is not our goal to eliminate all discrimination in software. Instead, it is our goal to empower the developers and stakeholders to identify and reason about discrimination in software. As described above, there are plenty of real-world examples in which companies would prefer to have discovered discrimination earlier, prior to release. Specifically, our technique's job is to identify if software discriminates with respect to a specific set of characteristics. If the stakeholder expects cat vs. dog discrimination, she would exclude it from the list of input characteristics to test. However, learning that the software frequently misclassifies black cats can help the stakeholder improve the software. Knowing if there is discrimination can lead to better-informed decision making.

There are two main challenges to measuring discrimination via testing. First, generating a practical set of test inputs sufficient for measuring discrimination, and second, processing those test inputs' executions to compute discrimination. This paper tackles both challenges, but the main contribution is computing discrimination from a set of executions. We are aware of no prior testing technique, neither automated nor manual, that produces a measure of a software system's *causal* discrimination. The paper also contributes within the space of efficient test input generation for the specific purpose of discrimination testing (see Section 4), but some prior work, specifically in combinatorial testing, e.g., [6, 44, 47, 80], may further help the efficiency of test generation, though these techniques have not been previously applied to discrimination testing. We leave a detailed examination of how combinatorial testing and other automated test generation can help further improve Themis to future work.

This paper's main contributions are:

(1) Formal definitions of software fairness and discrimination, including a causality-based improvement on the state-of-the-art definition of algorithmic fairness.
(2) Themis, a technique and open-source implementation — https://github.com/LASER-UMASS/Themis — for measuring discrimination in software.
(3) A formal analysis of the theoretical foundation of Themis, including proofs of monotonicity of discrimination that lead to provably sound two-to-three orders of magnitude improvements in test suite size, a proof of the relationship between fairness definitions, and a proof that Themis is more efficient on systems that exhibit more discrimination.
(4) An evaluation of the fairness of 20 real-world software instances (based on 8 software systems), 12 of which were designed with fairness in mind, demonstrating that (i) even when fairness is a design goal, developers can easily introduce discrimination in software, and (ii) Themis is an effective fairness testing tool.

Themis requires a schema for generating inputs, but does not require an oracle. Our causal fairness definition is designed specifically to be testable, unlike definitions that require probabilistic estimation or knowledge of the future [38]. Software testing offers a unique opportunity to conduct causal experiments to determine statistical causality [67]: One can run the software on an input (e.g., a defendant's criminal record), modify a specific input characteristic (e.g., the defendant's race), and observe if that modification *causes* a change in the output. We define software to be causally fair with respect to input characteristic $\chi$ if for all inputs, varying the value of $\chi$ does not alter the output. For example, a sentence-recommendation system is fair with respect to race if there are no two individuals who differ only in race but for whom the system's sentence recommendations differs. In addition to capturing causality, this definition requires no oracle — the equivalence of the output for the two inputs is itself the oracle — which helps fully automate test generation.

The rest of this paper is structured as follows. Section 2 provides an intuition to fairness measures and Section 3 formally defines software fairness. Section 4 describes Themis, our approach to fairness testing. Section 5 evaluates Themis. Finally, Section 6 places our work in the context of related research and Section 7 summarizes our contributions.

## 2 SOFTWARE FAIRNESS MEASURES

Suppose a bank employs LOAN software to decide if loan applicants should be given loans. LOAN inputs are each applicant's name, age, race, income, savings, employment status, and requested loan amount, and the output is a binary "give loan" or "do not give loan". For simplicity, suppose age and race are binary, with age either <40 or >40, and race either green or purple.

Some prior work on measuring and removing discrimination from algorithms [18, 19, 39–42, 88, 89, 91] has focused on what we call *group discrimination*, which says that to be fair with respect to an input characteristic, the distribution of outputs for each group should be similar. For example, the LOAN software is fair with respect to age if it gives loans to the same fractions of applicants <40 and >40. To be fair with respect to multiple characteristics, for example, age and race, all groups with respect to those characteristics — purple <40, purple >40, green <40, and green >40 — should have the same outcome fractions. The Calders-Verwer (CV) score [19] measures the strength of group discrimination as the difference between the largest and the smallest outcome fractions; if 30% of people <40 get the loan, and 40% of people >40 get the loan, then LOAN is 40% − 30% = 10% group discriminating.

While group discrimination is easy to reason about and measure, it has two inherent limitations. First, group discrimination may fail to observe some discrimination. For example, suppose that LOAN produces different outputs for two loan applications that differ in race, but are otherwise identical. While LOAN clearly discriminates with respect to race, the group discrimination score will be 0 if LOAN discriminates in the opposite way for another pair of applications. Second, software may circumvent discrimination detection. For example, suppose LOAN recommends loans for a random 30% of the purple applicants, and the 30% of the green applicants who have the most savings. Then the group discrimination score with respect to race will deem LOAN perfectly fair, despite a clear discrepancy in how the applications are processed based on race.

To address these issues, we define a new measure of discrimination. Software testing enables a unique opportunity to conduct *causal experiments* to determine statistical causation [67] between inputs and outputs. For example, it is possible to execute LOAN on two individuals identical in every way except race, and verify if changing the race causes a change in the output. *Causal discrimination* says that to be fair with respect to a set of characteristics, the software must produce the same output for every two individuals who differ only in those characteristics. For example, the LOAN software is fair with respect to age and race if for all pairs of individuals with identical name, income, savings, employment status, and requested loan amount but different race or age characteristics, LOAN either gives all of them or none of them the loan. The fraction of inputs for which software causally discriminates is a measure of causal discrimination.

Thus far, we have discussed software operating on the full input domain, e.g., every possible loan application. However, applying software to partial input domains may mask or effect discrimination. For example, while software may discriminate on some loan applications, a bank may care about whether that software discriminates only with respect to applications representative of their customers, as opposed to all possible human beings. In this case, a partial input domain may mask discrimination. If a partial input domain exhibits correlation between input characteristics, it can effect discrimination. For example, suppose older individuals have, on average, higher incomes and larger savings. If LOAN only considers income and savings in making its decision, even though it does not consider age, for this population, LOAN gives loans to a higher fraction of older individuals than younger ones. We call the measurement of group or causal discrimination on a partial input domain *apparent discrimination*. Apparent discrimination depends on the operational profile of the system system's use [7, 58]. Apparent discrimination is important to measure. For example, Amazon.com, Inc. software that determined where to offer free same-day delivery did not explicitly consider race but made race-correlated decisions because of correlations between race and other input characteristics [36, 52]. Despite the algorithm not looking at race explicitly, Amazon.com, Inc. would have preferred to have tested for this kind of discrimination.

## 3 FORMAL FAIRNESS DEFINITIONS

We make two simplifying assumptions. First, we define software as a black box that maps *input* characteristics to an *output* characteristic. While software is, in general, more complex, for the purposes of fairness testing, without loss of generality, this definition is sufficient: All user actions and environmental variables are modeled as input characteristics, and each software effect is modeled as an output characteristic. When software has multiple output characteristics, we define fairness with respect to each output characteristic separately. The definitions can be extended to include multiple output characteristics without significant conceptual reformulation. Second, we assume that the input characteristics and the output characteristic are categorical variables, each having a set of possible values (e.g., race, gender, eye color, age ranges, income ranges). This assumption simplifies our measure of causality. While our definitions do not apply directly to non-categorical input and output characteristics (such as continuous variables, e.g., int and double), they, and our techniques, can be applied to software with non-categorical input and output characteristics by using binning (e.g., age<40 and age>40). The output domain distance function (Definition 3.3) illustrates one way our definitions can be extended to continuous variables. Future work will extend our discrimination measures directly to a broader class of data types.

A *characteristic* is a categorical variable. An *input type* is a set of characteristics, an *input* is a valuation of an input type (assignment of a value to each characteristic), and an output is a single characteristic.

*Definition 3.1 (Characteristic).* Let $L$ be a set of value labels. A characteristic $\chi$ over $L$ is a variable that can take on the values in $L$.

*Definition 3.2 (Input type and input).* For all $n \in \mathbb{N}$, let $L_1, L_2, \ldots, L_n$ be sets of value labels. Then an input type $X$ over those value labels is a sequence of characteristics $X = \langle \chi_1, \chi_2, \ldots, \chi_n \rangle$, where for all $i \leq n$, $\chi_i$ is a characteristic over $L_i$.

An input of type $X$ is $k = \langle l_1 \in L_1, l_2 \in L_2, \ldots, l_n \in L_n \rangle$, a valuation of an input type.

We say the sizes of input $k$ and of input type $X$ are $n$.

Discrimination can be measured in software that makes decisions. When the output characteristic is binary (e.g., "give loan" vs. "do not give loan") the significance of the two different output values is clear. When outputs are not binary, identifying potential discrimination requires understanding the significance of differences in the output. For example, if the software outputs an ordering of hotel listings (that may be influenced by the computer you are using, as was the case when software used by Orbitz led Apple users to higher-priced hotels [53]), domain expertise is needed to compare two outputs and decide the degree to which their difference is significant. The *output domain distance function* encodes this expertise, mapping pairs of output values to a distance measure.

*Definition 3.3 (Output domain distance function).* Let $L_o$ be a set of value labels. Then for all $l_{o1}, l_{o2} \in L_o$, the output distance function is $\delta : L_o \times L_o \to [0..1]$ such that $l_{o1} = l_{o2} \implies \delta(l_{o1}, l_{o2}) = 0$.

The output domain distance function generalizes our work beyond binary outputs. For simplicity of exposition, for the remainder of this paper, we assume software outputs binary decisions — a natural domain for fairness testing. While true or false outputs (corresponding to decisions such as "give loan" vs. "do not give loan") are easier to understand, the output domain distance function enables comparing non-binary outputs in two ways. First, a *threshold* output domain distance function can determine when two outputs are dissimilar enough to warrant potential discrimination. Second, a *relational* output domain distance function can describe how different two inputs are and how much they contribute to potential discrimination. Definitions 3.5, 3.6, 3.8, and 3.7, could be extended to handle non-binary outputs by changing their exact output comparisons to fractional similarity comparisons using an output domain distance function, similar to the way inputs have been handled in prior work [24].

*Definition 3.4 (Decision software).* Let $n \in \mathbb{N}$ be an input size, let $L_1, L_2, \ldots, L_n$ be sets of value labels, let $X = \langle \chi_1, \chi_2, \ldots, \chi_n \rangle$ be an input type, and let $K$ be the set of all possible inputs of type $X$. Decision software is a function $\mathbb{S}: K \to \{\text{true, false}\}$. That is, when software $\mathbb{S}$ is applied to an input $\langle l_1 \in L_1, l_2 \in L_2, \ldots, l_n \in L_n \rangle$, it produces true or false.

The *group discrimination score* varies from 0 to 1 and measures the difference between fractions of input groups that lead to the same output (e.g., the difference between the fraction of green and purple individuals who are given a loan). This definition is based on the CV score [19], which is limited to a binary input type or a binary partitioning of the input space. Our definition extends to the more broad categorical input types, reflecting the relative complexity of arbitrary decision software. The group discrimination score with respect to a set of input characteristics is the maximum frequency with which the software outputs true minus the minimum such frequency for the groups that only differ in those input characteristics. Because the CV score is limited to a single binary partitioning, that difference represents all the encoded discrimination information in that setting. In our more general setting with multiple non-binary characteristics, the score focuses on the range — difference between the maximum and minimum — as opposed to the distribution. One could consider measuring, say, the standard deviation of the distribution of frequencies instead, which would better measure deviation from a completely fair algorithm, as opposed to the maximal deviation for two extreme groups.

*Definition 3.5 (Univariate group discrimination score $\tilde{d}$).* Let $K$ be the set of all possible inputs of size $n \in \mathbb{N}$ of type $X = \langle \chi_1, \chi_2, \ldots, \chi_n \rangle$ over label values $L_1, L_2, \ldots, L_n$. Let software $\mathbb{S}: K \to \{\text{true, false}\}$.

For all $i \leq n$, fix one characteristic $\chi_i$. That is, let $m = |L_i|$ and for all $\hat{m} \leq m$, let $K_{\hat{m}}$ be the set of all inputs with $\chi_i = l_{\hat{m}}$. ($K_{\hat{m}}$ is the set of all inputs with the $\chi_i^{\text{th}}$ characteristic fixed to be $l_{\hat{m}}$.) Let $p_{\hat{m}}$ be the fraction of inputs $k \in K_{\hat{m}}$ such that $\mathbb{S}(k) = \text{true}$. And let $P = \langle p_1, p_2, \ldots, p_m \rangle$.

Then the univariate group discrimination score with respect to $\chi_i$, denoted $\tilde{d}_{\chi_i}(\mathbb{S})$, is $\max(P) - \min(P)$.

For example, consider LOAN software that decided to give loan to 23% of green individuals, and to 65% of purple individuals. When computing LOAN's group discrimination score with respect to race, $\tilde{d}_{\text{race}}(\text{LOAN}) = 0.65 - 0.23 = 0.42$.

The multivariate group discrimination score generalizes the univariate version to multiple input characteristics.

*Definition 3.6 (Multivariate group discrimination score $\tilde{d}$).* For all $\alpha, \beta, \ldots, \gamma \leq n$, fix the characteristics $\chi_\alpha, \chi_\beta, \ldots, \chi_\gamma$. That is, let $m_\alpha = |L_\alpha|, m_\beta = |L_\beta|, \ldots, m_\gamma = |L_\gamma|$, let $\hat{m}_\alpha \leq m_\alpha, \hat{m}_\beta \leq m_\beta, \ldots, \hat{m}_\gamma \leq m_\gamma$, and $m = m_\alpha \times m_\beta \times \cdots \times m_\gamma$, let $K_{\hat{m}_\alpha, \hat{m}_\beta, \ldots, \hat{m}_\gamma}$ be the set of all inputs with $\chi_\alpha = l_{\hat{m}_\alpha}, \chi_\beta = l_{\hat{m}_\beta}, \ldots, \chi_\gamma = l_{\hat{m}_\gamma}$. ($K_{\hat{m}_\alpha, \hat{m}_\beta, \ldots, \hat{m}_\gamma}$ is the set of all inputs with the $\chi_\alpha$ characteristic fixed to be $l_{\hat{m}_\alpha}$, $\chi_\beta$ characteristic fixed to be $l_{\hat{m}_\beta}$, and so on.) Let $p_{\hat{m}_\alpha, \hat{m}_\beta, \ldots, \hat{m}_\gamma}$ be the fraction of inputs $k \in K_{\hat{m}_\alpha, \hat{m}_\beta, \ldots, \hat{m}_\gamma}$ such that $\mathbb{S}(k) = \text{true}$. And let $P$ be an unordered sequence of all $p_{\hat{m}_\alpha, \hat{m}_\beta, \ldots, \hat{m}_\gamma}$.

Then the multivariate group discrimination score with respect to $\chi_\alpha, \chi_\beta, \ldots, \chi_\gamma$, denoted $\tilde{d}_{\chi_\alpha, \chi_\beta, \ldots, \chi_\gamma}(\mathbb{S})$ is $\max(P) - \min(P)$.

Our *causal discrimination score* is a stronger measure of discrimination, as it seeks out causality in software, measuring the fraction of inputs for which changing specific input characteristics causes the output to change [67]. The causal discrimination score identifies changing which characteristics directly affects the output. As a result, for example, while the group and apparent discrimination scores penalize software that gives loans to different fractions of individuals of different races, the causal discrimination score penalizes software that gives loans to individuals of one race but not to otherwise identical individuals of another race.

*Definition 3.7 (Multivariate causal discrimination score $\vec{d}$).* Let $K$ be the set of all possible inputs of size $n \in \mathbb{N}$ of type $X = \langle \chi_1, \chi_2, \ldots, \chi_n \rangle$ over label values $L_1, L_2, \ldots, L_n$. Let software $\mathbb{S}: K \to \{\text{true}, \text{false}\}$. For all $\alpha, \beta, \ldots, \gamma \leq n$, let $\chi_\alpha, \chi_\beta, \ldots, \chi_\gamma$ be input characteristics.

Then the causal discrimination score with respect to $\chi_\alpha, \chi_\beta, \ldots, \chi_\gamma$, denoted $\vec{d}_{\chi_\alpha, \chi_\beta, \ldots, \chi_\gamma}(\mathbb{S})$ is the fraction of inputs $k \in K$ such that there exists an input $k' \in K$ such that $k$ and $k'$ differ only in the input characteristics $\chi_\alpha, \chi_\beta, \ldots, \chi_\gamma$, and $\mathbb{S}(k) \neq \mathbb{S}(k')$. That is, the causal discrimination score with respect to $\chi_\alpha, \chi_\beta, \ldots, \chi_\gamma$ is the fraction of inputs for which changing at least one of those characteristics causes the output to change.

Thus far, we have measured discrimination of the full input domain, considering every possible input with every value of every characteristic. In practice, input domains may be partial. A company may, for example, care about whether software discriminates only with respect to their customers (recall Section 2). *Apparent discrimination* captures this notion, applying group or causal discrimination score measurement to a subset of the input domain, which can be described by an operational profile [7, 58].

*Definition 3.8 (Multivariate apparent discrimination score).* Let $\ddot{K} \subseteq K$ be a subset of the input domain to $\mathbb{S}$. Then the apparent group discrimination score is the group discrimination score applied to $\ddot{K}$, and the apparent causal discrimination score is the causal discrimination score applied to $\ddot{K}$ (as opposed to applied to the full $K$).

Having defined discrimination, we now define the problem of checking software for discrimination.

*Definition 3.9 (Discrimination checking problem).* Given an input type $X$, decision software $\mathbb{S}$ with input type $X$, and a threshold $0 \leq \theta \leq 1$, compute all $X' \subseteq X$ such that $\tilde{d}_{X'}(\mathbb{S}) \geq \theta$ or $\vec{d}_{X'}(\mathbb{S}) \geq \theta$.

## 4 THE THEMIS SOLUTION

This section describes Themis, our approach to efficient fairness testing. To use Themis, the user provides a software executable, a desired confidence level, an acceptable error bound, and an input schema describing the format of valid inputs. Themis can then be used in three ways:

(1) Themis generates a test suite to compute the software group or causal discrimination score for a particular set of characteristics. For example, one can use Themis to check if, and how much, a software system discriminates against race and age.
(2) Given a discrimination threshold, Themis generates a test suite to compute all sets of characteristics against which a software group or causally discriminates more than that threshold.
(3) Given a manually-written or automatically-generated test suite, or an operational profile describing an input distribution [7, 58], Themis computes the apparent group or causal discrimination score for a particular set of characteristics. For example, one can use Themis to check if a system discriminates against race on a specific population of inputs representative of the way the

**Algorithm 1:** Computing group discrimination. Given a software $\mathbb{S}$ and a subset of its input characteristics $X'$, GROUPDISCRIMINATION returns $\tilde{d}_{X'}(\mathbb{S})$, the group discrimination score with respect to $X'$, with confidence $conf$ and error margin $\epsilon$.

GROUPDISCRIMINATION($\mathbb{S}$, $X'$, $conf$, $\epsilon$)
1  $minGroup \leftarrow \infty$, $maxGroup \leftarrow 0$, $testSuite \leftarrow \emptyset$ ▷*Initialization*
2  **foreach** $A$, where $A$ is a value assignment for $X'$ **do**
3      $r \leftarrow 0$ ▷*Initialize number of samples*
4      $count \leftarrow 0$ ▷*Initialize number of positive outputs*
5      **while** $r < \text{MAX\_SAMPLES}$ **do**
6          $r \leftarrow r + 1$
7          $k \leftarrow NewRandomInput(X' \leftarrow A)$ ▷*New input $k$ with $k.X' = A$*
8          $testSuite \leftarrow testSuite \cup \{k\}$ ▷*Add input to the test suite*
9          **if** $notCached(k)$ **then** ▷*No cached executions of $k$ exist*
10             $Compute(\mathbb{S}(k))$ ▷*Evaluate software on input $k$*
11             $CacheResult(k, \mathbb{S}(k))$ ▷*Cache the result*
12         **else** ▷*Retrieve cached result*
13             $\mathbb{S}(k) \leftarrow RetrieveCached(k)$
14         **if** $\mathbb{S}(k)$ **then**
15             $count \leftarrow count + 1$
16         **if** $r > \text{SAMPLING\_THRESHOLD}$ **then**
            ▷*After sufficient samples, check error margin*
17             $p \leftarrow \frac{count}{r}$ ▷*Current proportion of positive outputs*
18             **if** $conf.zValue\sqrt{\frac{p(1-p)}{r}} < \epsilon$ **then**
19                 **break** ▷*Achieved error $< \epsilon$, with confidence $conf$*
20     $maxGroup \leftarrow \max(maxGroup, p)$
21     $minGroup \leftarrow \min(minGroup, p)$
22 **return** $testSuite$, $\tilde{d}_{X'}(\mathbb{S}) \leftarrow maxGroup - minGroup$

system will be used. This method does not compute the score's confidence as it is only as strong as the developers' confidence that test suite or operational profile is representative of real-world executions.

Measuring group and causal discrimination exactly requires exhaustive testing, which is infeasible for nontrivial software. Solving the discrimination checking problem (Definition 3.9) further requires measuring discrimination over all possible subsets of characteristics to find those that exceed a certain discrimination threshold.

Themis addresses these challenges by employing three optimizations: (1) test caching, (2) adaptive, confidence-driven sampling, and (3) sound pruning. All three techniques reduce the number of test cases needed to compute both group and causal discrimination. Section 4.1 describes how Themis employs caching and sampling, and Section 4.2 describes how Themis prunes the test suite search space.

### 4.1 Caching and Approximation

GROUPDISCRIMINATION (Algorithm 1) and CAUSALDISCRIMINATION (Algorithm 2) present the Themis computation of multivariate group and causal discrimination scores with respect to a set of characteristics. These algorithms implement Definitions 3.6 and 3.7, respectively, and rely on two optimizations. We first describe these optimizations and then the algorithms.

**Test caching.** Precisely computing the group and causal discrimination scores requires executing a large set of tests. However, a lot of this computation is repetitive: tests relevant to group discrimination are also relevant to causal discrimination, and tests relevant to one set of characteristics can also be relevant to another set. This redundancy in fairness testing allows Themis to exploit caching to reuse test results without re-executing tests. *Test caching* has low storage overhead and offers significant runtime gains.

**Adaptive, confidence-driven sampling.** Since exhaustive testing is infeasible, Themis computes approximate group and causal discrimination scores through sampling. Sampling in Themis is adaptive, using the ongoing score computation to determine if a specified margin of error $\epsilon$ with a desired confidence level *conf* has been reached. Themis generates inputs uniformly at random using an input schema, and maintains the proportion of samples ($p$) for which the software outputs true (in GroupDiscrimination) or for which the software changes its output (in CausalDiscrimination). The margin of error for $p$ is then computed as:

$$error = z^* \sqrt{\frac{p(1-p)}{r}}$$

where $r$ is the number of samples so far and $z^*$ is the normal distribution $z^*$ score for the desired confidence level. Themis returns if $error < \epsilon$, or generates another test otherwise.

GroupDiscrimination (Algorithm 1) measures the group discrimination score with respect to a subset of its input characteristics $X'$. As per Definition 3.6, GroupDiscrimination fixes $X'$ to particular values (line 2) to compute what portion of all tests with $X'$ values fixed produce a true output. The while loop (line 5) generates random input assignments for the remaining input characteristics (line 7), stores them in the test suite, and measures the *count* of positive outputs. The algorithm executes the test, if that execution is not already cached (line 9); otherwise, the algorithm retrieves the software output from the cache (line 12). After passing the minimum sampling threshold (line 16), it checks if $\epsilon$ error margin is achieved with the desired confidence (line 18). If it is, GroupDiscrimination terminates the computation for the current group and updates the max and min values (lines 20–21).

CausalDiscrimination (Algorithm 2) similarly applies test caching and adaptive sampling. It takes a random test $k_0$ (line 4) and tests if changing any of its $X'$ characteristics changes the output. If $k_0$ result is not cached (line 6), the algorithm executes it and caches the result. It then iterates through tests $k$ that differ from $k_0$ in one or more characteristics in $X'$ (line 11). All generated inputs are stored in the test suite. The algorithm typically only needs to examine a small number of tests before discovering causal discrimination for the particular input (line 18). In the end, CausalDiscrimination returns the proportion of tests for which the algorithm found causal discrimination (line 25).

### 4.2 Sound Pruning

Measuring software discrimination (Definition 3.9) involves executing GroupDiscrimination and CausalDiscrimination over each subset of the input characteristics. The number of these executions grows exponentially with the number of characteristics. Themis relies on a powerful pruning optimization to dramatically reduce the

---

**Algorithm 2:** Computing causal discrimination. Given a software $\mathbb{S}$ and a subset of its input characteristics $X'$, CausalDiscrimination returns $\vec{d}_{X'}(\mathbb{S})$, the causal discrimination score with respect to $X'$, with confidence *conf* and error margin $\epsilon$.

CausalDiscrimination($\mathbb{S}, X', conf, \epsilon$)
1. $count \leftarrow 0; r \leftarrow 0, testSuite \leftarrow \emptyset$  ▷ *Initialization*
2. **while** $r <$ MAX_SAMPLES **do**
3.    $r \leftarrow r + 1$
4.    $k_0 \leftarrow$ NewRandomInput  ▷ *New input without value restrictions*
5.    $testSuite \leftarrow testSuite \cup \{k_0\}$  ▷ *Add input to the test suite*
6.    **if** $notCached(k_0)$ **then**  ▷ *No cached executions of $k_0$ exist*
7.       Compute($\mathbb{S}(k_0)$)  ▷ *Evaluate software on input $k_0$*
8.       CacheResult($k_0, \mathbb{S}(k_0)$)  ▷ *Cache the result*
9.    **else**  ▷ *Retrieve cached result*
10.       $\mathbb{S}(k_0) \leftarrow$ RetrieveCached($k_0$)
11.    **foreach** $k \in \{k \mid k \neq k_0; \forall \chi \notin X', k.\chi = k_0.\chi\}$ **do**
        ▷ *All inputs that match $k_0$ in every characteristic $\chi \notin X'$*
12.       $testSuite \leftarrow testSuite \cup \{k\}$  ▷ *Add input to the test suite*
13.       **if** $notCached(k)$ **then**  ▷ *No cached executions of $k$ exist*
14.          Compute($\mathbb{S}(k)$)  ▷ *Evaluate software on input $k$*
15.          CacheResult($k, \mathbb{S}(k)$)  ▷ *Cache the result*
16.       **else**  ▷ *Retrieve cached result*
17.          $\mathbb{S}(k) \leftarrow$ RetrieveCached($k$)
18.       **if** $\mathbb{S}(k) \neq \mathbb{S}(k_0)$ **then**  ▷ *Causal discrimination*
19.          $count = count + 1$
20.          **break**
21.    **if** $r >$ SAMPLING_THRESHOLD **then**
        ▷ *Once we have sufficient samples, check error margin*
22.       $p \leftarrow \frac{count}{r}$  ▷ *Current proportion of positive outputs*
23.       **if** $conf.zValue\sqrt{\frac{p(1-p)}{r}} < \epsilon$ **then**
24.          **break**  ▷ *Achieved error $< \epsilon$, with confidence conf*
25. **return** $testSuite, \vec{d}_{X'}(\mathbb{S}) \leftarrow p$

---

number of evaluated characteristics subsets. Pruning is based on a fundamental monotonicity property of group and causal discrimination: if a software $\mathbb{S}$ discriminates over threshold $\theta$ with respect to a set of characteristics $X'$, then $\mathbb{S}$ also discriminates over threshold $\theta$ with respect to all superset of $X'$. Once Themis discovers that $\mathbb{S}$ discriminates against $X'$, it can prune testing all supersets of $X'$.

Next, we formally prove group (Theorem 4.1) and causal (Theorem 4.2) discrimination monotonicity. These results guarantee that Themis pruning strategy is sound. DiscriminationSearch (Algorithm 3) uses pruning in solving the discrimination checking problem. As Section 5.4 will evaluate empirically, pruning leads to, on average, a two-to-three order of magnitude reduction in test suite size.

**Theorem 4.1 (Group discrimination monotonicity).** *Let $X$ be an input type and let $\mathbb{S}$ be a decision software with input type $X$. Then for all sets of characteristics $X', X'' \subseteq X, X'' \supseteq X' \implies \tilde{d}_{X''}(\mathbb{S}) \geq \tilde{d}_{X'}(\mathbb{S})$.*

**Proof.** Let $\tilde{d}_{X'}(\mathbb{S}) = \theta'$. Recall (Definition 3.6) that to compute $\tilde{d}_{X'}(\mathbb{S})$, we partition the space of all inputs into equivalence classes

**Algorithm 3:** Discrimination search. Given a software $\mathbb{S}$ with input type $X$ and a discrimination threshold $\theta$, DiscriminationSearch identifies all minimal subsets of characteristics $X' \subseteq X$ such that the (group or causal) discrimination score of $\mathbb{S}$ with respect to $X'$ is greater than $\theta$ with confidence $conf$ and error margin $\epsilon$.

DiscriminationSearch($\mathbb{S}$, $\theta$, $conf$, $\epsilon$)

1  $\mathbb{D} \leftarrow \emptyset$  ▷ *Initialize discriminating subsets of characteristics*
2  **for** $i = 1 \ldots |X|$ **do**
3     **foreach** $X' \subseteq X$, $|X'| = i$ **do**  ▷ *Check subsets of size i*
4        $discriminates \leftarrow$ false
5        **for** $X'' \in \mathbb{D}$ **do**
6           **if** $X'' \subseteq X'$ **then**
               ▷ *Supersets of a discriminating set discriminate (Theorems 4.1 and 4.2)*
7              $discriminates \leftarrow$ true
8              **break**
9        **if** $discriminates$ **then**
10          **continue**  ▷ *Do not store $X'$; $\mathbb{D}$ already contains a subset*
11       $\{testSuite, d\}$ = Discrimination($\mathbb{S}$, $X'$, $conf$, $\epsilon$)
         ▷ *Test $\mathbb{S}$ for (group or causal) discrimination with respect to $X'$*
12       **if** $d > \theta$ **then**  ▷ *$X'$ is a minimal discriminating set*
13          $\mathbb{D}.append(X')$

such that all elements in each equivalence class have identical value labels assigned to each characteristic in $X'$, then, compute the frequencies with which inputs in each equivalence class lead $\mathbb{S}$ to a true output, and finally compute the difference between the minimum and maximum of these frequencies. Let $\hat{p}'$ and $\check{p}'$ be those maximum and minimum frequencies, and $\hat{K}'$ and $\check{K}'$ be the corresponding equivalence classes of inputs.

Now consider the computation of $\theta'' = \tilde{d}_{X''}(\mathbb{S})$. Note that the equivalence classes of inputs for this computations will be strict subsets of the equivalence classes in the $\theta'$ computation. In particular, the equivalence subset $\hat{K}'$ will be split into several equivalence classes, which we call $\hat{K}''_1, \hat{K}''_2, \ldots$ There are two possibilities: (1) either the frequency with which the inputs in each of these subclasses lead $\mathbb{S}$ to a true output equal the frequency of $\hat{K}'$, or (2) some subclasses have lower frequencies and some have higher than $\hat{K}'$ (since when combined, they must equal that of $\hat{K}'$). Either way, the maximum frequency of the $\hat{K}''_1, \hat{K}''_2, \ldots, \hat{K}''_j$ subclasses is $\geq \hat{K}'$. And therefore, the maximum overall frequency $\hat{p}''$ for all the equivalence classes in the computation of $\theta''$ is $\geq \hat{p}'$. By the same argument, the minimum overall frequency $\check{p}''$ for all the equivalence classes in the computation of $\theta''$ is $\leq \check{p}'$. Therefore, $\theta'' = (\hat{p}'' - \check{p}'') \geq (\hat{p}' - \check{p}') \leq = \theta'$, and therefore, $X'' \supseteq X' \implies \vec{d}_{X''}(\mathbb{S}) \geq \vec{d}_{X'}(\mathbb{S})$. □

THEOREM 4.2 (CAUSAL DISCRIMINATION MONOTONICITY). *Let $X$ be an input type and let $\mathbb{S}$ be a decision software with input type $X$. Then for all sets of characteristics $X', X'' \subseteq X, X'' \supseteq X' \implies \vec{d}_{X''}(\mathbb{S}) \geq \vec{d}_{X'}(\mathbb{S})$.*

PROOF. Recall (Definition 3.7) that the causal discrimination score with respect to $X'$ is the fraction of inputs for which changing the value of at least one characteristic in $X'$ changes the output. Consider $K'$, the entire set of such inputs for $X'$, and similarly $K''$, the entire set of such inputs for $X''$. Since $X'' \supseteq X'$, every input in $K'$ must also be in $K''$ because if changing at least one characteristic in $X'$ changes the output and those characteristics are also in $X''$. Therefore, the fraction of such inputs must be no smaller for $X''$ than for $X'$, and therefore, $X'' \supseteq X' \implies \vec{d}_{X''}(\mathbb{S}) \geq \vec{d}_{X'}(\mathbb{S})$. □

A further opportunity for pruning comes from the relationship between group and causal discrimination. As Theorem 4.3 shows, if software group discriminates against a set of characteristics, it must causally discriminate against that set at least as much.

THEOREM 4.3. *Let $X$ be an input type and let $\mathbb{S}$ be a decision software with input type $X$. Then for all sets of characteristics $X' \subseteq X$, $\tilde{d}_{X'}(\mathbb{S}) \leq \vec{d}_{X'}(\mathbb{S})$.*

PROOF. Let $\tilde{d}_{X'}(\mathbb{S}) = \theta$. Recall (Definition 3.6) that to compute $\tilde{d}_{X'}(\mathbb{S})$, we partition the space of all inputs into equivalence classes such that all elements in each equivalence class have identical value labels assigned to each characteristic in $X'$. It is evident that same equivalence class inputs have same values for characteristics in $X'$ and the ones in different equivalence classes differ in at least one of the characteristics in $X'$.

Now, $\tilde{d}_{X'}(\mathbb{S}) = \theta$ means that for $\theta$ fraction of inputs, the output is true, and after changing just some values of $X'$ (producing an input in another equivalence class), the output is false. This is because if there were $\theta' < \theta$ fraction of inputs with a different output when changing the equivalence classes, then $\tilde{d}_{X'}(\mathbb{S})$ would have been $\theta'$. Hence $\vec{d}_{X'}(\mathbb{S}) > \theta$. □

## 5 EVALUATION

In evaluating Themis, we focused on two research questions:

**RQ1:** Does research on discrimination-aware algorithm design (e.g., [18, 40, 88, 91]) produce fair algorithms?
**RQ2:** How effective do the optimizations from Section 4 make Themis at identifying discrimination in software?

To answer these research questions, we carried out three experiments on twenty instances of eight software systems that make financial decisions.[1] Seven of the eight systems (seventeen out of the twenty instances) are written by original system developers; we reimplemented one of the systems (three instances) because original source code was not available. Eight of these software instances use standard machine learning algorithms to infer models from datasets of financial and demographic data. These systems make no attempt to avoid discrimination. The other twelve instances are taken from related work on devising discrimination-aware algorithms [18, 40, 88, 91]. These software instances use the same datasets and attempt to infer discrimination-free solutions. Four of them focus on not discriminating against race, and eight against gender. Section 5.1 describes our subject systems and the two datasets they use.

---

[1]We use the term system instance to mean instantiation of a software system with a configuration, using a specific dataset. Two instances of the same system using different configurations and different data are likely to differ significantly in their behavior and in their discrimination profile.

## 5.1 Subject Software Systems

Our twenty subject instances use two financial datasets. The Adult dataset (also known as the Census Income dataset)[2] contains financial and demographic data for 45K individuals; each individual is described by 14 attributes, such as occupation, number of work hours, capital gains and losses, education level, gender, race, marital status, age, country of birth, income, etc. This dataset is well vetted: it has been used by others to devise discrimination-free algorithms [18, 88, 89], as well as for non-discrimination purposes [3, 43, 86]. The Statlog German Credit dataset[3] contains credit data for 1,000 individuals, classifying each individual as having "good" or "bad" credit, and including 20 other pieces of data for each individual, such as gender, housing arrangement, credit history, years employed, credit amount, etc. This dataset is also well vetted: it has been used by others to devise discrimination-free algorithms [18, 89], as well as for non-discrimination purposes [25, 31].

We use a three-parameter naming scheme to refer to our software instances. An example of an instance name is $\text{Ⓐ}_{\text{race}}^{\text{census}}$. The "**A**" refers to the system used to generate the instance (described next). The "census" refers to the dataset used for the system instance. This value can be "census" for the Adult Census Income dataset or "credit" for the Statlog German Credit dataset. Finally, the "race" refers to a characteristic the software instance attempts to not discriminate against. In our evaluation, this value can be "race" or "gender" for the census dataset and can only be "gender" for the credit dataset.[4] Some of the systems make no attempt to avoid discrimination and their names leave this part of the label blank.

Prior research has attempted to build discrimination-free systems using these two datasets [18, 40, 88, 91]. We contacted the authors and obtained the source code for three of these four systems, Ⓐ [88], Ⓒ [18], and Ⓓ [91], and reimplemented the other, Ⓑ [40], ourselves. We verified that our reimplementation produced results consistent with the the evaluation of the original system [40]. We additionally used standard machine learning libraries as four more discrimination-unaware software systems Ⓔ, Ⓕ, Ⓖ, and Ⓗ, on these datasets. We used scikit-learn [68] for Ⓔ (naive Bayes), Ⓖ (logistic regression), and Ⓗ (support vector machines), and a publicly available decision tree implementation [90] for Ⓕ and for our reimplementation of Ⓑ.

The four discrimination-free software systems use different methods to attempt to reduce or eliminate discrimination:

Ⓐ is a modified logistic regression approach that constrains the regression's loss function with the covariance of the characteristics' distributions that the algorithm is asked to be fair with respect to [88].

Ⓑ is a modified decision tree approach that constrains the splitting criteria by the output characteristic (as the standard decision tree approach does) and also the characteristics that the algorithm is asked to be fair with respect to [40].

Ⓒ manipulates the training dataset for a naive Bayes classifier. The approach balances the dataset to equate the number of inputs that lead to each output value, and then tweaks the dataset by introducing noise of flipping outputs for some inputs, and by introducing weights for each datapoint [18].

Ⓓ is a modified decision tree approach that balances the training dataset to equate the number of inputs that lead to each output value, removes and repeats some inputs, and flips output values of inputs close to the decision tree's decision boundaries, introducing noise around the critical boundaries [91].

**Configuring the subject systems.** The income characteristic of the census dataset is binary, representing the income being above or below $50,000. Most prior research developed systems that use the other characteristics to predict the income; we did the same. For the credit dataset, the systems predict if the individual's credit is "good" or "bad". We trained each system, separately on the census and credit datasets. Thus, for example, the $\text{Ⓖ}^{\text{credit}}$ instance is the logistic-regression-based system trained on the credit dataset. For the census dataset, we randomly sampled 15.5K individuals to balance the number who make more than and less than $50,000, and trained each system on the sampled subset using 13 characteristics to classify each individual as either having above or below $50,000 income. For the credit dataset, we similarly randomly sampled 600 individuals to balance the number with "good" and "bad" credit, and trained each system on the sampled subset using the 20 characteristics to classify each individual's credit as "good" or "bad".

Each discrimination-aware system can be trained to avoid discrimination against sensitive characteristics. In accord with the prior work on building these systems [18, 40, 88, 91], we chose gender and race as sensitive characteristics. Using all configurations exactly as described in the prior work, we created 3 instances of each discrimination-aware system. For example, for system Ⓐ, we have $\text{Ⓐ}_{\text{gender}}^{\text{census}}$ and $\text{Ⓐ}_{\text{race}}^{\text{census}}$, two instances trained on the census data to avoid discrimination on gender and race, respectively, and $\text{Ⓐ}_{\text{gender}}^{\text{credit}}$, an instance trained on the credit data to avoid discrimination on gender. The left column of Figure 1 lists the twenty system instances we use as subjects.

## 5.2 Race and Gender Discrimination

We used Themis to measure the group and causal discrimination scores for our twenty software instances with respect to race and, separately, with respect to gender. Figure 1 presents the results. We make the following observations:

- **Themis is effective.** Themis is able to (1) verify that many of the software instances do not discriminate against race and gender, and (2) identify the software that does.
- **Discrimination is present even in instances designed to avoid discrimination.** For example, a discrimination-aware decision tree approach trained not to discriminate against gender, $\text{Ⓑ}_{\text{gender}}^{\text{census}}$, had a causal discrimination score over 11%: more than 11% of the individuals had the output flipped just by altering the individual's gender.
- **The causal discrimination score detected critical evidence of discrimination missed by the group score.** Often, the group and causal discrimination scores conveyed the same information, but there were cases in which Themis detected causal discrimination even though the group discrimination score was low. For example, for (B)$_{\text{gender}}^{\text{census}}$, the causal score was more than 21× higher than the group score (11.27% vs. 0.52%).

---

[2]https://archive.ics.uci.edu/ml/datasets/Adult
[3]https://archive.ics.uci.edu/ml/datasets/Statlog+(German+Credit+Data)
[4]The credit dataset combines marital status and gender into a single characteristic; we refer to it simply as gender in this paper for exposition. The credit dataset does not include race.

| System Instance | Race group | Race causal | Gender group | Gender causal |
|---|---|---|---|---|
| (A) credit/gender | – | – | 3.78% | 3.98% |
| (A) census/gender | 2.10% | 2.25% | 3.80% | 3.80% |
| (A) census/race | 2.10% | 1.13% | 8.90% | 7.20% |
| (B) credit/gender | – | – | 0.80% | 2.30% |
| (B) census/gender | 36.55% | 38.40% | 0.52% | 11.27% |
| (B) census/race | 2.28% | 1.78% | 5.84% | 5.80% |
| (C) credit/gender | – | – | 0.35% | 0.18% |
| (C) census/gender | 2.43% | 2.91% | < 0.01% | < 0.01% |
| (C) census/race | 0.08% | 0.08% | 35.20% | 34.50% |
| (D) credit/gender | – | – | 0.23% | 0.29% |
| (D) census/gender | 0.21% | 0.26% | < 0.01% | < 0.01% |
| (D) census/race | 0.12% | 0.13% | 4.64% | 4.94% |
| (E) credit | – | – | 0.32% | 0.37% |
| (E) census | 0.74% | 0.85% | 0.26% | 0.32% |
| (F) credit | – | – | 0.05% | 0.06% |
| (F) census | 0.11% | 0.05% | < 0.01% | < 0.01% |
| (G) credit | – | – | 3.94% | 2.41% |
| (G) census | 0.02% | 2.80% | < 0.01% | < 0.01% |
| (H) credit | – | – | < 0.01% | < 0.01% |
| (H) census | < 0.01% | 0.01% | < 0.01% | < 0.01% |

Figure 1: The group and causal discrimination scores with respect to race and gender. Some numbers are missing because the credit dataset does not contain information on race.

- **Today's discrimination-aware approaches are insufficient.** The (B) approach was designed to avoid a variant of group discrimination (as are other discrimination-aware approaches), but this design is, at least in some conditions, insufficient to prevent causal discrimination. Further, focusing on avoiding discriminating against one characteristic may create discrimination against another, e.g., (B)$_{\text{gender}}^{\text{census}}$ limits discrimination against gender but discriminates against race with a causal score of 38.40%.
- **There is no clear evidence that discrimination-aware methods outperform discrimination-unaware ones.** In fact, the discrimination-unaware approaches typically discriminated less than their discrimination-aware counterparts, with the exception of logistic regression.

### 5.3 Computing Discriminated-Against Characteristics

To evaluate how effective Themis is at computing the discrimination checking problem (Definition 3.9), we used Themis to compute the sets of characteristics each of the twenty software instances discriminates against causally. For each instance, we first used a threshold of 75% to find all subsets of characteristics against which the instance discriminated. We next examined the discrimination with respect to each of the characteristics in those sets individually. Finally, we checked the causal discrimination scores for pairs of those characteristics that are sensitive, as defined by prior work [18, 40, 88, 91] (e.g., race, age, marital status, etc.). For example, if Themis found that an instance discriminated causally against {capital gains, race, marital status}, we checked the causal discrimination score for {capital gains}, {race}, {marital status}, and then {race, marital status}. Figure 2 reports which sensitive characteristics each instance discriminates against by at least 5%.

| | | | |
|---|---|---|---|
| (A)$_{\text{race}}^{\text{census}}$ | $\vec{d}_{\{g,r\}} = 13.7\%$ | (C)$_{\text{gender}}^{\text{census}}$ | $\vec{d}_{\{m\}} = 35.2\%$ |
| (B)$_{\text{gender}}^{\text{census}}$ | $\vec{d}_{\{g,m,r\}} = 77.2\%$ | (D)$_{\text{gender}}^{\text{census}}$ | $\vec{d}_{\{m\}} = 12.9\%$ |
| | $\vec{d}_{\{g,r\}} = 52.5\%$ | | $\vec{d}_{\{c\}} = 7.6\%$ |
| | $\vec{d}_{\{g\}} = 11.2\%$ | (D)$_{\text{race}}^{\text{census}}$ | $\vec{d}_{\{m\}} = 16.2\%$ |
| | $\vec{d}_{\{m\}} = 36.1\%$ | | $\vec{d}_{\{m,r\}} = 52.3\%$ |
| | $\vec{d}_{\{r\}} = 36.6\%$ | (E)$^{\text{census}}$ | $\vec{d}_{\{m\}} = 7.9\%$ |
| (B)$_{\text{race}}^{\text{census}}$ | $\vec{d}_{\{g\}} = 5.8\%$ | (F)$^{\text{census}}$ | $\vec{d}_{\{c,r\}} = 98.1\%$ |
| (C)$_{\text{gender}}^{\text{credit}}$ | $\vec{d}_{\{a\}} = 7.6\%$ | | $\vec{d}_{\{r,e\}} = 76.3\%$ |
| (C)$_{\text{race}}^{\text{census}}$ | $\vec{d}_{\{a\}} = 25.9\%$ | (G)$^{\text{census}}$ | $\vec{d}_{\{e\}} = 14.8\%$ |
| | $\vec{d}_{\{g\}} = 35.2\%$ | | |
| | $\vec{d}_{\{g,r\}} = 41.5\%$ | | |

Figure 2: Sets of sensitive characteristics that the subject instances discriminate against causally at least 5% and that contribute to subsets of characteristics that are discriminated against at least 75%. We abbreviate sensitive characteristics as: (a)ge, (c)ountry, (g)ender, (m)arital status, (r)ace, and r(e)lation.

Themis was able to discover significant discrimination. For example, (B)$_{\text{gender}}^{\text{census}}$ discriminates against gender, marital status, and race with a causal score of 77.2%. That means for 77.2% of the individuals, changing only the gender, marital status, or race causes the output of the algorithm to flip. Even worse, (F)$^{\text{census}}$ discriminates against country and race with a causal score of 98.1%.

It is possible to build an algorithm that appears to be fair with respect to a characteristic in general, but discriminates heavily against that characteristic when the input space is partitioned by another characteristic. For example, an algorithm may give the same fraction of white and black individuals loans, but discriminate against black Canadian individuals as compared to white Canadian individuals. This is the case with (B)$_{\text{gender}}^{\text{census}}$, for example, as its causal discrimination score against gender is 11.2%, but against gender, marital status, and race is 77.2%. Prior work on fairness has not considered this phenomenon, and these findings suggest that the software designed to produce fair results sometimes achieved fairness at the global scale by creating severe discrimination for certain groups of inputs.

This experiment demonstrates that Themis effectively discovers discrimination and can test software for unexpected software discrimination effects across a wide variety of input partitions.

### 5.4 Themis Efficiency and the Pruning Effect

Themis uses pruning to minimize test suites (Section 4.2). We evaluated the efficiency improvement due to pruning by comparing the number of test cases needed to achieve the same confidence and error bound with and without pruning. Figure 3 shows the number of test cases needed for each of the twenty software instances to achieve a confidence level of 99% and 0.05 error bound, with and without pruning. Pruning reduces the number of test cases by, on average, a factor of 2,849 for group and 148 for causal discrimination.

| | | Group | | Causal | |
|---|---|---|---|---|---|
| Ⓐ credit/gender | $\frac{934,230}{167,420} =$ | 5.6× | $\frac{29,623}{15,764} =$ | 1.9× |
| Ⓐ census/race | $\frac{7,033,150}{672} =$ | 10,466 × | $\frac{215,000}{457} =$ | 470 × |
| Ⓑ credit/gender | $\frac{934,230}{3,413} =$ | 274 × | $\frac{29,623}{1,636} =$ | 18 × |
| Ⓑ census/race | $\frac{6,230,000}{80,100} =$ | 78 × | $\frac{20,500}{6,300} =$ | 33 × |
| Ⓒ credit/gender | $\frac{934,230}{113} =$ | 8,268 × | $\frac{29,623}{72} =$ | 411 × |
| Ⓒ census/race | $\frac{7,730,120}{75,140} =$ | 103 × | $\frac{235,625}{6,600} =$ | 36 × |
| Ⓓ credit/gender | $\frac{934,230}{720} =$ | 1,298 × | $\frac{29,623}{472} =$ | 63 × |
| Ⓓ census/race | $\frac{7,730,120}{6,600,462} =$ | 1.2× | $\frac{235,625}{200,528} =$ | 1.2× |
| Ⓔ credit | $\frac{934,230}{145} =$ | 6,443 × | $\frac{29,623}{82} =$ | 361 × |
| Ⓔ census | $\frac{7,730,120}{84,040} =$ | 92 × | $\frac{235,625}{7,900} =$ | 30 × |
| Ⓕ credit | $\frac{934,230}{3,410} =$ | 274 × | $\frac{29,623}{2,647} =$ | 11 × |
| Ⓕ census | $\frac{6,123,000}{461} =$ | 13,282 × | $\frac{205,000}{279} =$ | 735 × |
| Ⓖ credit | $\frac{934,230}{187} =$ | 4,996 × | $\frac{29,623}{152} =$ | 195 × |
| Ⓖ census | $\frac{7,730,120}{5,160,125} =$ | 1.5× | $\frac{235,625}{190,725} =$ | 1.2× |
| Ⓗ credit | $\frac{934,230}{412,020} =$ | 2.3× | $\frac{29,623}{10,140} =$ | 2.9× |
| Ⓗ census | $\frac{1,530,000}{1,213,500} =$ | 1.3× | $\frac{510,000}{324,582} =$ | 1.6× |
| arithmetic mean | | 2,849 × | | 148 × |
| geometric mean | | 151 × | | 26.4× |

Figure 3: Pruning greatly reduces the number of tests needed to compute both group and causal discrimination. We present here the computation that is needed for the experiment of Figure 2: finding all subsets of characteristics for which the software instances discriminate with a score of at least 75%, for a 99% confidence and error margin 0.05. For each technique, we show the number of tests needed without pruning divided by the number of tests needed with pruning, and the resulting factor reduction in the number of tests. For example, reducing the number of tests needed to compute the group discrimination score from 7,033,150 to 672 (2$^{nd}$ row) is an improvement of a factor of 10,466.

The more a system discriminates, the more effective pruning is, making Themis more efficient because pruning happens when small sets of characteristics discriminate above the chosen threshold. Such sets enable pruning away larger supersets of characteristics. Theorem 5.1 formalizes this statement.

THEOREM 5.1 (PRUNING MONOTONICITY). *Let X be an input type and $\mathbb{S}$ and $\mathbb{S}'$ be decision software with input types X. If for all $X' \subseteq X$, $\vec{d}_{X'}(\mathbb{S}) \geq \vec{d}_{X'}(\mathbb{S}')$ (respectively, $\tilde{d}_{X'}(\mathbb{S}) \geq \tilde{d}_{X'}(\mathbb{S}')$), then for all $X'' \subseteq X$, if Themis can prune $X''$ when computing DISCRIMINATIONSEARCH($\mathbb{S}'$, $\theta$, conf, $\epsilon$), then it can also prune $X''$ when computing DISCRIMINATIONSEARCH($\mathbb{S}$, $\theta$, conf, $\epsilon$).*

PROOF. For Themis to prune $X''$ when computing DISCRIMINATIONSEARCH($\mathbb{S}'$, $\theta$, conf, $\epsilon$), there must exist a set $\hat{X}'' \subsetneq X''$ such that $\vec{d}_{\hat{X}''}(\mathbb{S}') \geq \theta$. Since $\vec{d}_{\hat{X}''}(\mathbb{S}) \geq \vec{d}_{\hat{X}''}(\mathbb{S}') \geq \theta$, when computing

DISCRIMINATIONSEARCH($\mathbb{S}$, $\theta$, conf), Themis can also prune $X''$. The same argument holds for group discrimination $\tilde{d}$. □

We measured this effect by measuring pruning while decreasing the discrimination threshold $\theta$; decreasing $\theta$ effectively simulates increasing system discrimination. We verified that pruning increased when $\theta$ decreased (or equivalently, when discrimination increased). For example, Themis needed 3,413 tests to find sets of characteristics that Ⓑ$^{credit}_{gender}$ discriminated with a score of more than $\theta = 0.7$, but only 10 tests when we reduced $\theta$ to 0.6. Similarly, the number of tests for Ⓕ$^{credit}$ dropped from 920 to 10 when lowering $\theta$ from 0.6 to 0.5. This confirms that Themis is more efficient when the benefits of fairness testing increase because the software discriminates more.

### 5.5 Discussion

In answering our two research questions, we found that (1) State-of-the-art approaches for designing fair systems often miss discrimination and Themis can detect such discrimination via fairness testing. (2) Themis is effective at finding both group and causal discrimination. While we did not evaluate this directly, Themis can also measure apparent discrimination (Definition 3.8) via a developer-provided test suite or operational profile. (3) Themis employs provably sound pruning to reduce test suite size and becomes more effective for systems that discriminate more. Overall, pruning reduced test suite sizes, on average, two to three orders of magnitude.

## 6 RELATED WORK

Software discrimination is a growing concern. Discrimination shows up in many software applications, e.g., advertisements [75], hotel bookings [53], and image search [45]. Yet software is entering domains in which discrimination could result in serious negative consequences, including criminal justice [5, 28], finance [62], and hiring [71]. Software discrimination may occur unintentionally, e.g., as a result of implementation bugs, as an unintended property of self-organizing systems [11, 13, 15, 16], as an emergent property of component interaction [12, 14, 17, 49], or as an automatically learned property from biased data [18, 19, 39–42, 88, 89, 91].

Some prior work on measuring fairness in machine learning classifiers has focused on the Calders-Verwer (CV) score [19] to measure discrimination [18, 19, 39–42, 87–89, 91]. Our group discrimination score generalizes the CV score to the software domain with more complex inputs. Our causal discrimination score goes beyond prior work by measuring causality [67]. An alternate definition of discrimination is that a "better" input is never deprived of the "better" output [24]. That definition requires a domain expert to create a distance function for comparing inputs; by contrast, our definitions are simpler, more generally applicable, and amenable to optimization techniques, such as pruning. Reducing discrimination (CV score) in classifiers [18, 40, 88, 91], as our evaluation has shown, often fails to remove causal discrimination and discrimination against certain groups. By contrast, our work does not attempt to remove discrimination but offers developers a tool to identify and measure discrimination, a critical first step in removing it. Problem-specific discrimination measures, e.g., the contextual bandits problem, have demonstrated that fairness may result in otherwise suboptimal behavior [38]. By contrast, our work is general and we believe that striving for fairness may be a principal requirement in a system's design.

Counterfactual fairness requires output probability distributions to match for input populations that differ only in the label value of a sensitive input characteristic [51]. Counterfactual fairness is related to causal fairness but can miss some instances of discrimination, e.g., if LOAN shows preferential treatment *for* some purple inputs, but at the same time *against* some other similar purple inputs.

FairTest, an implementation of the unwarranted associations framework [79] uses manually written tests to measure four kinds of discrimination scores: the CV score and a related ratio, mutual information, Pearson correlation, and a regression between the output and sensitive inputs. By contrast, our approach generates tests automatically and measures causal discrimination.

Causal testing computes pairs of similar inputs whose outputs differ. However, input characteristics may correlate, e.g., education correlates with age, so perturbing some characteristics without perturbing others may create inputs not representative of the real world. FairML [1] uses orthogonal projection to co-perturb characteristics, which can mask some discrimination, but find discrimination that is more likely to be observed in real-world scenarios, somewhat analogously to our apparent discrimination measure.

Combinatorial testing minimizes the number of tests needed to explore certain combinations of input characteristics. For example, all-pairs testing generates tests that evaluate every possible value combination for every pair of input characteristics, which can be particularly helpful when testing software product lines [6, 44, 46, 47]. The number of tests needed to evaluate every possible value pair can be significantly smaller than the exhaustive testing alternative since each test can simultaneously contribute to multiple value pairs [22, 44, 80]. Such combinatorial testing optimizations are complementary to our work on discrimination testing. Our main goal is to develop a method to process test executions to measure software discrimination, whereas that is not a goal of combinatorial testing. Advances in combinatorial testing, e.g., using static or dynamic analyses for vacuity testing [8, 33] or to identify configuration options that cannot affect a test's output [48], can directly improve efficiency of discrimination testing by identifying that changing a particular input characteristic cannot affect a particular test's output, and thus no causal discrimination is possible with respect to that particular input. We leave such optimizations to future work.

It is possible to test for discrimination software without explicit access to it. For example, AdFisher [23] collects information on how changes in Google ad settings and prior visited webpages affect the ads Google serves. AdFisher computes a variant of group discrimination, but it could be integrated with Themis and its algorithms to measure causal discrimination.

Themis measures apparent discrimination by either executing a provided test suite, or by generating a test suite following a provided operational profile. Operational profiles [58] describe the input distributions likely to be observed in the field. Because developer-written test suites are often not as representative of field executions as developers would like [81], operational profiles can significantly improve the effectiveness of testing by more accurately representing real-world system use [7, 50] and the use of operational profiles has been shown to more accurately measure system properties, such as reliability [35]. The work on operational profiles is complementary to ours: Themis uses operational profiles and work on more efficient test generation from operational profiles can directly benefit discrimination testing. Meanwhile no prior work on operational profile testing has measured software discrimination.

Causal relationships in data management systems [54, 55] can help explain query results [57] and debug errors [82–84] by tracking and using data provenance [56]. For software systems that use data management, such provenance-based reasoning may aid testing for causal relationships between input characteristics and outputs. Our prior work on testing software that relies on data management systems has focused on data errors [59, 60], whereas this work focuses on testing fairness.

Automated testing research has produced tools to generate tests, including random testing, such as Randoop [63, 64], NightHawk [4], JCrasher [20], CarFast [65], and T3 [69]; search-based testing, such as EvoSuite [29], TestFul [9], and eToc [78]; dynamic symbolic execution tools, such as DSC [37], Symbolic PathFinder [66], jCUTE [70], Seeker [76], Symstra [85], and Pex [77], among others; and commercial tools, such as Agitar [2]. The goal of the generated tests is typically finding bugs [29] or generating specifications [21]. These tools deal with more complex input spaces than Themis, but none of them focus on testing fairness and they require oracles whereas Themis does not need oracles as it measures discrimination by comparing tests' outputs. Future work could extend these tools to generate fairness tests, modifying test generation to produce pairs of inputs that differ only in the input characteristics being tested. While prior work has tackled the oracle problem [10, 26, 30] typically using inferred pre- and post-conditions or documentation, our oracle is more precise and easier to compute, but is only applicable to fairness testing.

## 7 CONTRIBUTIONS

We have formally defined software fairness testing and introduced a causality-based measure of discrimination. We have further described Themis, an approach and its open-source implementation for measuring discrimination in software and for generating efficient test suites to perform these measurements. Our evaluation demonstrates that discrimination in software is common, even when fairness is an explicit design goal, and that fairness testing is critical to measuring discrimination. Further, we formally prove soundness of our approach and show that Themis effectively measures discriminations and produces efficient test suites to do so. With the current use of software in society-critical ways, fairness testing research is becoming paramount, and our work presents an important first step in merging testing techniques with software fairness requirements.

## ACKNOWLEDGMENT

This work is supported by the National Science Foundation under grants no. CCF-1453474, IIS-1453543, and CNS-1744471.

Fairness Testing: Testing Software for Discrimination                                    ESEC/FSE'17, September 4–8, 2017, Paderborn, Germany